\documentclass{jkas}

\def\year{2018} 
\def\month{February} 
\def\volume{51} 
\def\beginpage{17} 
\def\endpage{25} 
\def\received{June 26, 2017} 
\def\accepted{February 6, 2018} 
\date{Received \received; accepted \accepted}

\usepackage{flushend} 
\newcommand\ion[2]{{#1}\,{\sc #2}} 
\usepackage[ps2pdf, breaklinks, colorlinks, citecolor=blue, linkcolor=blue, menucolor=blue, urlcolor=blue]{hyperref}

\title{Search for Exoplanets around Northern Circumpolar Stars \\ III. Long-Period Radial Velocity Variations\\ in HD 18438 and HD 158996}

\author[1,4]{Tae-Yang~Bang}
\author[2,3]{Byeong-Cheol~Lee}
\author[2,3]{Gwang-Hui~Jeong}
\author[2,3]{Inwoo~Han}
\author[1,4]{Myeong-Gu~Park}

\affil[1]{Department of Astronomy and Atmospheric Sciences, Kyungpook National University, 80 Daehakro, Bukgu, Daegu 41566, Korea; \email{qkdxodid1230@knu.ac.kr, mgp@knu.ac.kr}}
\affil[2]{Korea Astronomy and Space Science Institute, 776 Daedukdae-ro,
  Yuseong-gu, Daejeon 34055, Korea; \email{bclee@kasi.re.kr, tlotv@kasi.re.kr, iwhan@kasi.re.kr}}
\affil[3]{Korea University of Science and Technology, 217 Gajeong-ro, Yuseong-gu, Daejeon 34113, Korea}
\affil[4]{Research and Training Team for Future Creative Astrophysicists and Cosmologists (BK21 Plus Program)}

\def\corrauthor{
M.-G. Park
}

\def\runningauthor{
Bang et al.
}

\def\runningtitle{
Analysis of Long-Period Radial Velocity Variations in HD 18438 and HD 158996
}

\def\keywords{
stars: individual: HD 18438, HD 158996 --- techniques: radial velocities
}

\def\abstracttext{
Detecting exoplanets around giant stars sheds light on the later-stage evolution of planetary systems. We observed the M giant HD 18438 and the K giant HD 158996 as part of a Search for Exoplanets around Northern circumpolar Stars (SENS) and obtained 38 and 24 spectra from 2010 to 2017 using the high-resolution Bohyunsan Observatory Echelle Spectrograph (BOES) at the 1.8m telescope of Bohyunsan Optical Astronomy Observatory in Korea. We obtained precise RV measurements from the spectra and found long-period radial velocity (RV) variations with period 719.0 days for HD 18438 and 820.2 days for HD 158996. We checked the chromospheric activities using \ion{Ca}{ii} H and H$_{\alpha}$ lines, $\emph{HIPPARCOS}$ photometry and line bisectors to identify the origin of the observed RV variations. In the case of HD 18438, we conclude that the observed RV variations with period 719.0 days are likely to be caused by the pulsations because the periods of $\emph{HIPPARCOS}$ photometric and H$_{\alpha}$ EW variations for HD 18438 are similar to that of RV variations in Lomb-Scargle periodogram, and there are no correlations between bisectors and RV measurements. In the case of HD 158996, on the other hand, we did not find any similarity in the respective periodograms nor any correlation between RV variations and line bisector variations. In addition, the probability that the real rotational period can be as longer than the RV period for HD 158996 is only about 4.3\%. Thus we conclude that observed RV variations with a period of 820.2 days of HD 158996 are caused by a planetary companion, which has the minimum mass of 14.0 $\it M_\mathrm{Jup}$, the semi-major axis of 2.1 AU, and eccentricity of 0.13 assuming the stellar mass of 1.8 $M_{\odot}$. HD 158996 is so far one of the brightest and largest stars to harbor an exoplanet candidate.
}

\begin{document}
\jkashead 


\section{Introduction\label{sec:intro}}

Since the first discovery of an exoplanet with the radial velocity (RV) method in 1995 (\citealt{Mayor}), about 18\% of all known exoplanets have been detected with precise RV methods so far. Host stars of these systems are mostly late-type main sequence (MS) stars because they have many narrow spectral lines suitable for precise RV measurements.

Searches for exoplanets around late-type giants using the RV method are actively pursued (\citealt{Frink}; \citealt{Hatzes2005}; \citealt{dollinger}; \citealt{Sato07}; \citealt{Han2010}) also because planet-harboring giants can give information on how stellar evolution affects planets. In addition, they have sharp spectral lines, thus enabling us to measure RV shifts with high precision. However, giants have more sources contributing to RV variation than MS stars, such as stellar activity, rotational modulation of surface features, and stellar pulsations. Therefore more careful investigations of the origin of RV variations are needed.

To find exoplanets around giants, we have conducted the SENS program (\citealt{Lee15}) over 7 years. We selected 224 stars from the \emph{HIPPARCOS} catalog using the following criteria: 1) $\delta$ $\geq$ 70\,$^{\circ}$; 2) 5.0 $<$ \emph{m}$_{v}$ $<$ 7.0; 3) 0.6 $<$ \emph{B$-$V} $<$ 1.6; and 4) \emph{HIPPARCOS}$_{\rm scat}$ $<$ 0.05 magnitude. We found 25 of them to harbor planets ($\sim$ 11\%) so far. The planet occurrence rate is consistent with other studies, about 10 $\sim$ 15\% (\citealt{Johnson}; \citealt{Kraus}; \citealt{Howard}; \citealt{Fressin}).

In this paper, we analyze the RV variations of two giant stars, HD 18438 and HD 158996, selected from the SENS program based on their characteristics. HD 18438 and HD 158996 are among the bigger and brighter giants in our sample. In Section~\ref{sec:obs}, we describe the observations and data reduction. In Section~\ref{sec:stellar}, we present stellar parameters. The RV measurements, orbit diagrams, and phase diagrams are described in Section~\ref{sec:rvs}. We analyze the spectral lines and photometric data in order to identify the origins of RV variations in Section~\ref{sec:identify}. Section~\ref{sec:results} shows our results, and summary and discussion are provided in Section~\ref{sec:con}.

\begin{figure}[!t]
\centering
\includegraphics[trim=0mm 5mm 5mm 25mm, clip, width=75mm]{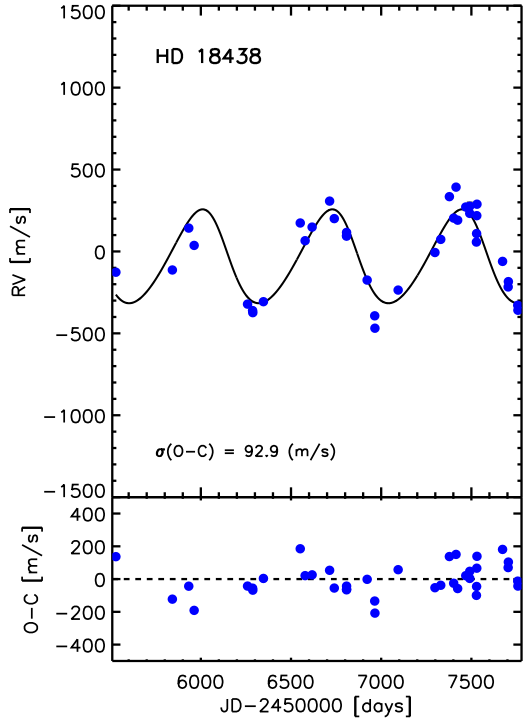}
\caption{RV measurements and Keplerian orbital fit (solid line) for HD 18438 with a period of 719.0 days.}.\label{fig:jkasfig1}
\end{figure}

\section{Observations and Reduction\label{sec:obs}}

We obtained all spectra using the high-resolution Bohyunsan Observatory Echelle Spectrograph (BOES; \citealt{Kim}) at the 1.8 meter telescope of Bohyunsan Optical Astronomy Observatory in Korea (BOAO). The BOES covers a wide wavelength range from 3500~{\AA} to 10\,500~{\AA} in one exposure. We used a fiber with a diameter of 200 ${\mu}$m, which provides a resolving power (R) of 45\,000. An iodine absorption (I$_{2}$) cell was used for precise wavelength calibration with a range from 4900~{\AA} to 6100~{\AA}. Our data reduction used the IRAF package, DECH (\citealt{Galaz}), the RVI2CELL code (\citealt{Han2007}) for RV measurements, and Systemic Console (\citealt{Meschi}) for analysis and fitting of RV measurements.

\begin{figure}[!t]
\centering
\includegraphics[trim=0mm 5mm 5mm 25mm, clip, width=75mm]{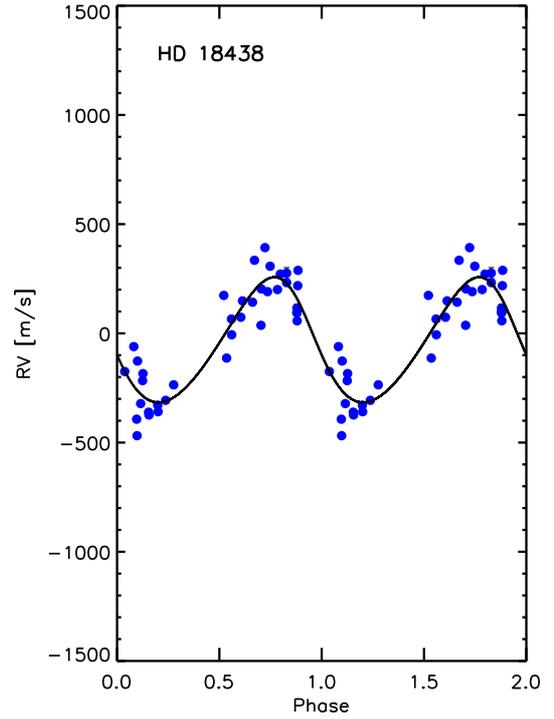}
\caption{The phase diagram of HD 18438 with a period of 719.0 days.}.\label{fig:jkasfig2}
\end{figure}

\begin{table*}[t!]
\caption{Stellar parameters for HD 18438 and HD 158996.\label{tab:jkastable1}}
\centering
\begin{tabular}{lcccc}
\toprule
Parameter & Unit  & HD 18438 & HD 158996 & Reference\\
\midrule
Spectral type					& 		    	 &M1 III		 & K5 III		 &\emph{HIPPARCOS} (\citealt{ESA})\\
$\textit{$m_{v}$}$							& mag		 &5.6 		 &5.8 		 &\cite{van}\\
$\emph{B$-$V}$ 							& mag 		 &1.7 		 &1.5		 &\cite{van}\\
$\pi$  						& mas 		 & $4.7\pm0.4$		 & 3.4$\pm$0.3 	 &\cite{van}\\
							&			 & $4.45\pm0.8$ & --- 	 &\cite{Gaia}\\
d 							&  pc	 	 & 224.7		 & 294.1 		 & This work\\
$\mathrm{[Fe/H]}$				& dex	    	 & $-0.4\pm0.1$ & $-0.2\pm0.1$ & This work\\
 $T_{\rm eff}$				&  K	   	 & 3844		 & 4165 		 &\cite{Mc}\\
							&			 & $4220\pm20$		 & $4069\pm30$		 &This work\\
log $\it g$					& cgs      	 & $0.9\pm0.1$  & $1.2\pm0.1$   &\cite{daSilva}\\\
							&			 & $2.2\pm0.1$	 & $1.4\pm0.1$	 & This work\\
$v_{\mathrm{micro}}$			& km s$^{-1}$ & $2.7\pm0.4$	 & $1.7\pm0.1$ & This work\\
RV 							& km s$^{-1}$ & $-37.6$ 	 & --- 		 	 &\cite{Willson}\\
							&			 &		--- 	 & $-8.7$ 		 &\cite{Gont}\\
$\textit{$L_{\star}$}$ 		& $L_{\odot}$  & 833.0		 & 1034.7	 	 &\cite{Anderson}\\
$\textit{$M_{\star}$}$			& $M_{\odot}$ & $1.1\pm0.1$ & $1.8\pm0.3$ 	 & This work\\
$\textit{$R_{\star}$}$ 		& $R_{\odot}$  & $60.7\pm6.6$ & $50.3\pm4.3$	 & This work\\
Age 							& Gyr	      & $6.1\pm2.0$ & $1.6\pm0.6$ & This work\\
 $v_{\mathrm{rot}}$ sin $i$		& km s$^{-1}$ & $5.5\pm0.2$ & $4.2\pm0.8$		 & This work\\
							&			 &	--- & $2.5\pm1.1$ &\cite{deMedeiros}\\
$P_{\mathrm{rot}}$/sin $i$ 		& days		 & 562		& 627		 & This work\\
\bottomrule
\end{tabular}
\end{table*}

\begin{table}[t!]
\centering
\setlength{\tabcolsep}{1.1mm}  
\caption{RV measurements for HD 18438 from November 2010 to
January 2017.}\label{tab:jkastable2}
\begin{tabular}{crr|crr}
\toprule
JD         & $\Delta$RV  & $\pm \sigma$ &        JD & $\Delta$RV  & $\pm \sigma$  \\
$-$2\,450\,000 & m\,s$^{-1}$ &  m\,s$^{-1}$ & $-$2\,450\,000  & m\,s$^{-1}$ &  m\,s$^{-1}$  \\
\midrule
5529.0770& $-$126.8& 9.4& 7094.0008& $-$235.8& 14.3\\
5842.2338& $-$113.2& 9.4& 7298.0154& $-$6.5 &11.7\\
5933.1005& 142.5 &10.9& 7330.1158& 73.6& 8.7\\
5962.9812&36.8 &10.8& 7378.1104& 334.5 &9.7\\
6259.127/& $-$321.6& 9.7& 7401.9251& 203.5& 10.8\\
6287.0339& $-$360.9& 12.9& 7414.9487& 392.6& 9.6\\
6288.1542& $-$373.7& 10.2& 7423.9433& 190.9 &10.1\\
6347.0366& $-$306.7& 9.2& 7468.9383 &271.3 &9.6\\
6551.2300& 173.7& 12.4& 7490.9975& 275.8 &25.3\\
6965.2620& $-$468.6& 9.3& 7758.0762& $-$358.9 &9.2\\
\bottomrule
\end{tabular}
\end{table}

\begin{table}[t!]
\setlength{\tabcolsep}{1.28mm}
\centering
\caption{RV measurements for HD 158996 from June 2010 to
January 2017.}\label{tab:jkastable3}
\begin{tabular}{crr|crr}
\toprule
JD         & $\Delta$RV  & $\pm \sigma$ &        JD & $\Delta$RV  & $\pm \sigma$  \\
$-$2\,450\,000 & m\,s$^{-1}$ &  m\,s$^{-1}$ & $-$2\,450\,000  & m\,s$^{-1}$ &  m\,s$^{-1}$  \\
\midrule
5357.0628& 51.5& 10.2 &7171.2045 &154.1 &12.3\\
5664.1472& $-28.8$& 9 &7298.9853 &35.7 &11.3\\
5843.0259& $-134.4$& 10 &7475.2480& $-176.1$ &9.3\\
6409.1197& 15.1 &8.6 &7525.1435 &$-192.9$ &9.3\\
6740.2149 &$-100.3$& 9.3 &7529.0204 &$-224.9$& 9.3\\
6805.2073 &$-6.4$ &9 &7530.1504 &$-247.7$ &9.2\\
6921.9780& $-19.9$ &10.4 &7531.0752& $-241.3$ &12.1\\
6970.8793& 46.3& 9.6 &7672.9253 &$-19.4$ &11.7\\
7066.2705 &207.6& 14.8 &7703.9122 &20.7& 9.4\\
7068.2330 &199.7 &10.5 &7704.9693 &64.3 &9.7\\
7094.1928 &168.4 &15.5 &7756.9184 &208.2 &9.3\\
7148.1184 &155.4 &11.1& 7757.9208 &65.1 &10.3\\
\bottomrule
\end{tabular}
\end{table}

\section{Stellar Model\label{sec:stellar}}

Compared to MS stars, giant stars have more complex RV variations because they often have a long-period stellar rotation with stellar activity such as spots, flares, and long-period pulsation. To confirm planetary companions with a long orbital period, calculating accurate stellar parameters is important for identification of various (non-exoplanet) origins of RV variations.

\subsection{Stellar Parameters\label{sec:parameters}}

We obtained the fundamental photometric parameters of HD 18438 and HD 158996 from the \emph{HIPPARCOS} catalog. Parallax of HD 18438 was taken from the Gaia DR 1 (\citealt{Gaia}). Stellar atmospheric parameters such as $\rm{T_{eff}}$, $\mathrm{[Fe/H]}$, $v_{\mathrm{micro}}$ and log $\it g$ were derived from TGVIT stellar model code (\citealt{Takeda05}). We used 147 and 227 equivalent widths (EW) of Fe I and Fe II lines of two target stars, respectively. However, because of the TGVIT is useful for stars with a surface temperature of over 4,000 K, stellar parameters of HD 18438 show some differences with values in the literature. We obtained stellar mass, radius, age, log $\it g$ using the online tool (\url{http://stevoapd.inaf.it/cgi-bin/param}; \citealt{daSilva}) which is based on a Bayesian estimation method. The stellar parameters of HD 18438 and HD 158996 are summarized in Table~\ref{tab:jkastable1}.

\subsection{Rotational Velocity and Period\label{sec:rot}}

Giants generally have rotation periods of several hundred days. Rotational modulation of surface features can cause long-term RV variations. Hence, the rotational period is a very important parameter to identify the origins of RV variations. If there is activity in the stellar chromosphere, RV variations may occur modulated by the rotation period. Rotational velocities ($v_{\mathrm{rot}}$ sin $i$) of the two target stars were estimated from SPTOOL code (\citealt{Takeda08}), which calculates the line-broadening by stellar rotation using five elements in the wavelength range from 6080~{\AA} to 6089~{\AA}. We estimated rotational velocities to be 5.5 $\pm$ 0.2 km\,s$^{-1}$ and 4.2 $\pm$ 0.8 km s$^{-1}$ for HD 18438 and HD 158996, respectively. The value for HD 158896 overlaps with $v_{\mathrm{rot}}$ sin $i$ = 2.5 $\pm$ 1.1 km s$^{-1}$ by \cite{deMedeiros}. These estimated rotational velocities are slightly faster than those of general giants. From our values we calculated the probability that the rotational period of the given star is longer than the observed RV period, under the assumption that the real $v_{\mathrm{rot}}$ sin $i$ and the stellar radius follow a normal distribution with mean and standard deviation equal to the observationally determined values, and that the rotational axis of the star is randomly oriented. The probability for HD 18438 to have a rotation period longer than the observed period of 719.0 days is 0.1\% and that for HD 158996 longer than the observed period of 820.2 days is 4.3\%. Hence, it is unlikely that the period seen in RV variation has originated from stellar rotation, neither for HD 18438 nor for HD 158996.

\begin{figure}[!t]
\centering
\includegraphics[trim=0mm 5mm 5mm 25mm, clip, width=75mm]{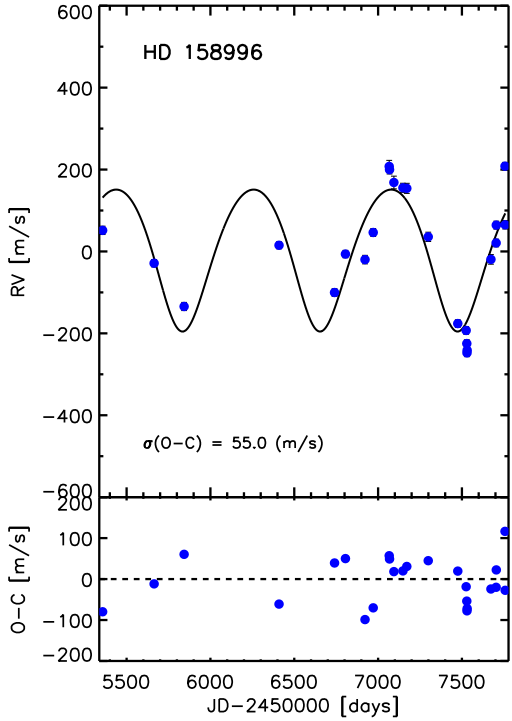}
\caption{RV measurements and Keplerian orbital fit (solid line) for HD 158996 with a period of 820.2 days.}.\label{fig:jkasfig3}
\end{figure}

\section{Radial Velocity Variations\label{sec:rvs}}

\begin{figure}[!t]
\centering
\includegraphics[trim=0mm 5mm 5mm 25mm, clip, width=75mm]{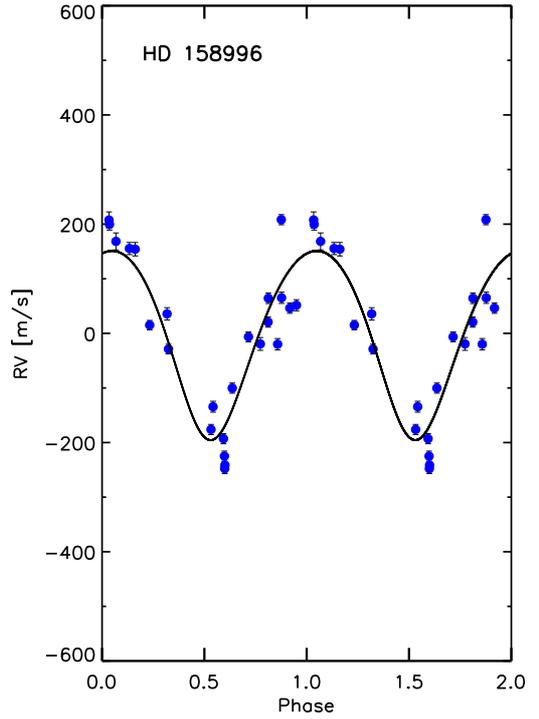}
\caption{The phase diagram of HD 158996 with a period of 820.2 days.}.\label{fig:jkasfig4}
\end{figure}

We took 38 spectra for HD 18438 from November 2010 to January 2017 and 24 spectra for HD 158996 from June 2010 to January 2017. We measured RVs from observed spectra using the RVI2CELL code. The RV measurements for HD 18438 and HD 158996 are given in Table \ref{tab:jkastable2} and \ref{tab:jkastable3}, respectively. The period was determined from Lomb-Scargle periodograms (Scargle 1982). Figure~\ref{fig:jkasfig1} shows the Keplerian fitting (curve) of RV measurements of HD 18438 with a variation period of 719.0 days, a semi-amplitude of 288.5 m\,s$^{-1}$, and an eccentricity of 0.12 (top panel) and residuals of 92.9 m\,s$^{-1}$ (bottom panel). Figure~\ref{fig:jkasfig2} shows the phase diagram of HD 18438. Figure~\ref{fig:jkasfig3} shows the Keplerian fitting of RV measurements of HD 158996 with a variation period of 820.2 days, a semi-amplitude of 207.0 m\,s$^{-1}$, and an eccentricity of 0.13 (top panel) and residuals of 57.8 m\,s$^{-1}$ (bottom panel). Figure~\ref{fig:jkasfig4} shows the phase diagram of HD 158996.

\section{Chromospheric Activity, Photometric and Line Profile Variations\label{sec:identify}}

Long-period RV variations can be produced not only by the orbiting companion but also by stellar pulsations or rotational modulation of surface features, possibly from chromospheric activity. Search for chromospheric activity or variations in photometry or in line shape can help to confirm or rule out each process. We calculated the variation periods of those parameters through the Lomb-Scargle periodograms. RVs, \emph{HIPPARCOS}, bisectors and H$_{\alpha}$ EW measurements (from top to bottom) of HD 18438 and HD 158996 are in Figures \ref{fig:jkasfig5} and \ref{fig:jkasfig6}, respectively.

\subsection{Chromospheric Activity\label{sec:activities}}

\begin{figure}[!t]
\centering
\includegraphics[trim=1mm 4mm 1mm 12mm, clip, width=80mm]{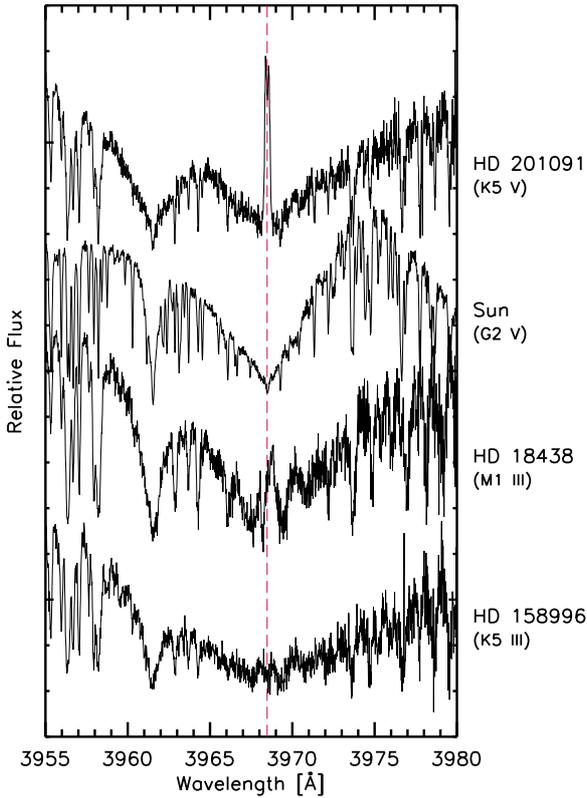}
\caption{Spectra of \ion{Ca}{ii} H lines for our sample stars, Sun, and chromospheric active star HD 201091. Vertical dashed line is the center of \ion{Ca}{ii} H line profiles. There are weak emission lines for HD 18438.}.\label{fig:jkasfig7}
\end{figure}

\begin{figure}[!t]
\centering
\includegraphics[trim=1mm 4mm 1mm 12mm, clip, width=80mm]{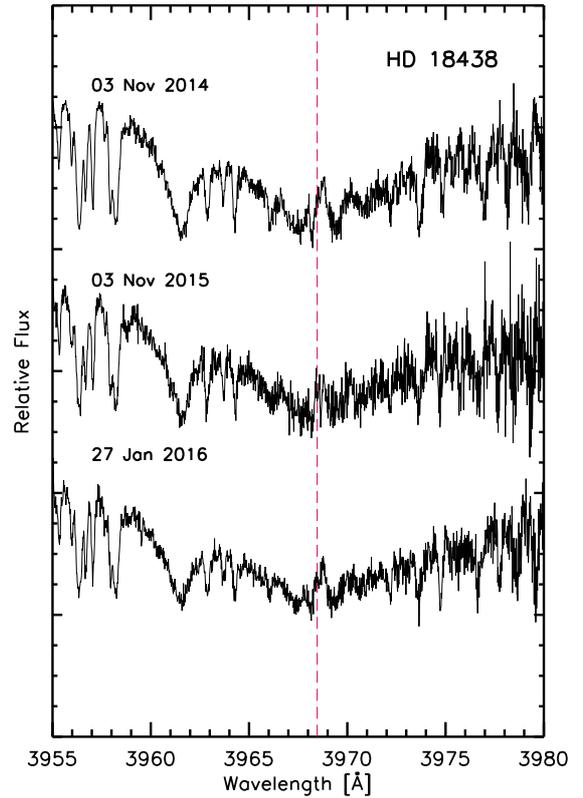}
\caption{The variations of \ion{Ca}{ii} H lines for HD 18438 at different phase. Averaged \ion{Ca}{ii} H line profiles at the minimum (top panel), around zero (middle panel) and maximum (bottom panel) part of the RV curve. There are weak emission lines at the center of \ion{Ca}{ii} H lines (vertical dashed line).}.\label{fig:jkasfig8}
\end{figure}

\begin{figure}[!t]
\centering
\includegraphics[trim=2mm 5mm 2mm 11mm, clip, width=80mm]{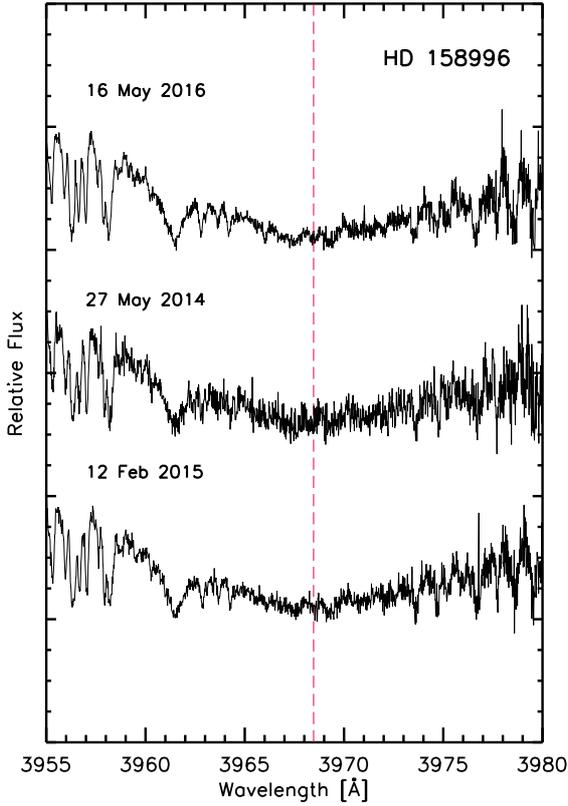}
\caption{The variations of \ion{Ca}{ii} H line for HD 158996 at different phase. Averaged \ion{Ca}{ii} H line profiles at the minimum (top panel), around zero (middle panel) and maximum (bottom panel) part of the RV curve. There are no emission features at the center of the \ion{Ca}{ii} H lines (vertical dashed line)}.\label{fig:jkasfig9}
\end{figure}

The \ion{Ca}{ii} H line was first discovered as an indicator of stellar activity by \cite{Eberhard}. Activity in the stellar chromosphere produce emission features at the \ion{Ca}{ii} H line center. Figure~\ref{fig:jkasfig7} shows the \ion{Ca}{ii} H lines for the chromospheric active star HD 201091, Sun, and our sample stars (from top to bottom). There is a weak emission feature near the center of \ion{Ca}{ii} H line for HD 18438. We examined the \ion{Ca}{ii} H line profiles at different RV phases to identify any systemic changes in stellar chromospheric activities. Figures \ref{fig:jkasfig8} and \ref{fig:jkasfig9} show \ion{Ca}{ii} H lines of HD 18438 and HD 158996 at different RV phases (from top to bottom), respectively. Figure~\ref{fig:jkasfig8} shows weak emission lines near the center of \ion{Ca}{ii} H lines of HD 18438. HD 158996, however, does not show such obvious features at the line center although one or two bumps are present near the center (Figure~\ref{fig:jkasfig9}). \cite{Sato10} shows the region of \ion{Ca}{ii} H line for chromospheric active star HD 120048, which shows a velocity scatter of about 20 m\,s$^{-1}$. Core reversal in \ion{Ca}{ii} H for HD 158996, if any, is much weaker than that of HD 120048. In addition, calculated amplitude of RV variations for HD 158996 is over 200 m\,s$^{-1}$ from Keplerian fitting and there are no systematic differences with respect to the different RV phase. It seems rather unlikely that the emission feature, if any, in HD 158996 implies strong enough chromospheric activity that can produce RV variations as large as 200 m\,s$^{-1}$.

\begin{figure}[!t]
\centering
\includegraphics[trim=0mm 2mm 5mm 20mm, clip, width=80mm]{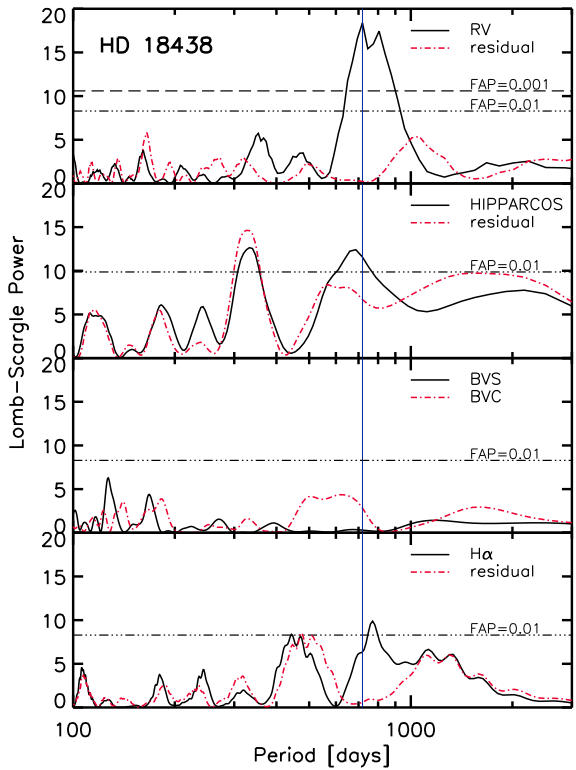}
\caption{Lomb-Scargle periodogram of RV measurements, \emph{HIPPARCOS}, bisectors and H$_{\alpha}$ EW measurements (from top to bottom) for HD 18438. Vertical solid line is the period of 719.0 days.}.\label{fig:jkasfig5}
\end{figure}

\begin{figure}[!t]
\centering
\includegraphics[trim=0mm 2mm 5mm 20mm, clip, width=80mm]{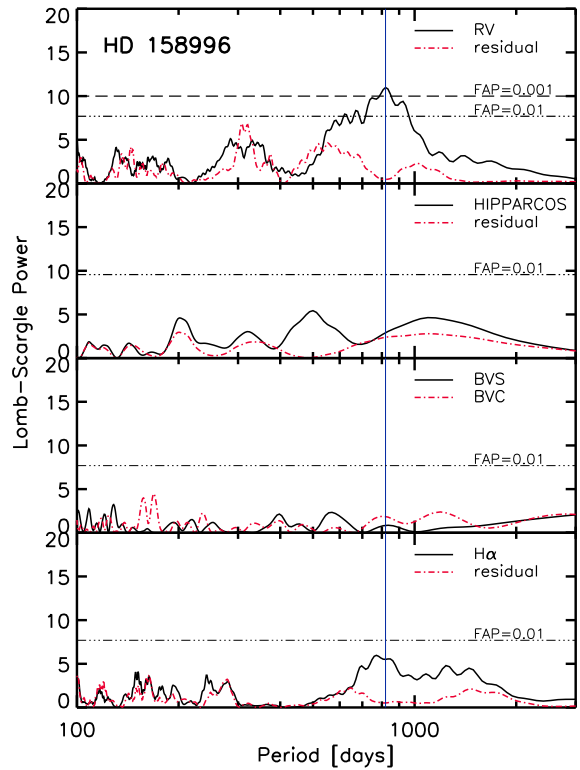}
\caption{Lomb-Scargle periodogram of RV measurements, \emph{HIPPARCOS}, bisectors and H$_{\alpha}$ EW measurements (from top to bottom) for HD 158996. Vertical solid line is the period of 820.2 days.}.\label{fig:jkasfig6}
\end{figure}

We checked EW variations of H$_{\alpha}$, another stellar activity indicator (\citealt{Montes}; \citealt{Kurster}; \citealt{Lee12}; \citealt{Hatzes2015}). The Lomb-Scargle periodogram of H$_{\alpha}$ EW variations in each star is shown in the bottom panel of Figures \ref{fig:jkasfig5} and \ref{fig:jkasfig6}. We found that a period of $\sim$ 783 days for HD 18438 and $\sim$ 775 days for HD 158996. The period of the significant peak of H$_{\alpha}$ EW variations in HD 18438 is similar to that of RV variations (bottom panel in Figure~\ref{fig:jkasfig5}). This suggests that there are stellar chromospheric activities in HD 18438, which have caused periodic RV variations. In the case of HD 158996, the peak of  H$_{\alpha}$ EW variations located near the RV period has FAP greater than 0.01 and is a part of a broad bump around the period of 1000 days. So the peak alone cannot prove or disprove the RV variations from chromospheric activity.

We therefore calculated the S index (\citealt{Baliunas}) for seven K-giant stars from our SENS program. S indices of these stars ranges from 0.000 to 0.006 with 0.002 for HD 158996, showing no correlation between S index and RV variation. Compared to other stars with similar S indices, HD 158996 has a much higher radial velocity variations. The mean value of maximum RV variations for seven K-giant stars is about 86 $\pm$ 55 m\,s$^{-1}$, which is much smaller than that of HD 158996, about 250 m\,s$^{-1}$ (This is not the amplitude K in Keplerian orbit fit). Thus we conclude that chromospheric activities, if any, can not produce the observed RV variations.

\subsection{HIPPARCOS Photometry\label{sec:hip}}

If the observed RV variations are from pulsations or rotations, which generally accompany photometric variations, the period of the photometric variations will overlap with the period of the RV variations. $\emph{HIPPARCOS}$ photometry data were analyzed. They were not contemporaneous with our RV measurements. We found that HD 158996 show several statistically insignificant peaks in the periodogram, none of which coincide with the RV period (Figure~\ref{fig:jkasfig6}). On the other hand, the Lomb-Scargle periodogram for HD 18438 shows a significant peak at $\sim$ 700 days, close to 719 day RV period (Figure~\ref{fig:jkasfig5}). In addition, there is another strong peak around 350 days, about a year, and it is half of period of a significant peak at $\sim$ 700 days. We checked window function of RV measurements and there is no similaritiy between the Lomb-Scargle periodograms of window function and RV measurements. Hence, it appears that the variations of RVs and $\emph{HIPPARCOS}$ photometric in HD 18438 are accompanied.

\subsection{Bisector Analysis\label{sec:bisector}}

Variable asymmetries in the spectral line profile might be caused by the rotation of a star with an inhomogeneous surface (\citealt{Hatzes1998}; \citealt{Queloz}). Bisector velocity span (BVS) is defined as the difference of RV measurements between the high and low flux levels of the line profile. Bisector velocity curvature is the difference in the velocity span of the upper half and the lower half of the line profile. These two parameters reflect any variations in the line shape.

We used mean values of five sharp absorption lines for HD 18438 (\ion{Fe}{i} 6085.2 {\AA}, \ion{Fe}{i} 6141.7 {\AA}, \ion{V}{i} 6251.8 {\AA}, \ion{Ti}{i} 6325.1 {\AA}, and \ion{Ca}{i} 6572.8 {\AA}) and HD 158996 (\ion{Fe}{i} 6085.2 {\AA}, \ion{Fe}{i} 6141.7 {\AA}, \ion{V}{i} 6251.8 {\AA}, \ion{Fe}{i} 6322.7 {\AA}, and \ion{Fe}{i} 6421.3 {\AA}) for bisector analysis. Figure~\ref{fig:jkasfig10} shows mean values of BVS (open diamonds) and BVC (filled circles) variations against RVs for HD 18438 and HD 158996. There are no correlations between RV measurements and bisectors in both stars. We conclude that there are no significant correlations between RV measurements and those parameters in HD 18438 and HD 158996.

The Lomb-Scargle periodograms of bisectors are shown in the third panel of Figures \ref{fig:jkasfig5} and \ref{fig:jkasfig6}, respectively. No significant peaks are found for both HD 18438 and HD 158996.

\section{Results\label{sec:results}}

\subsection{HD 18438}

\begin{figure}[!t]
\centering
\includegraphics[trim=5mm 5mm 5mm 25mm, clip, width=80mm]{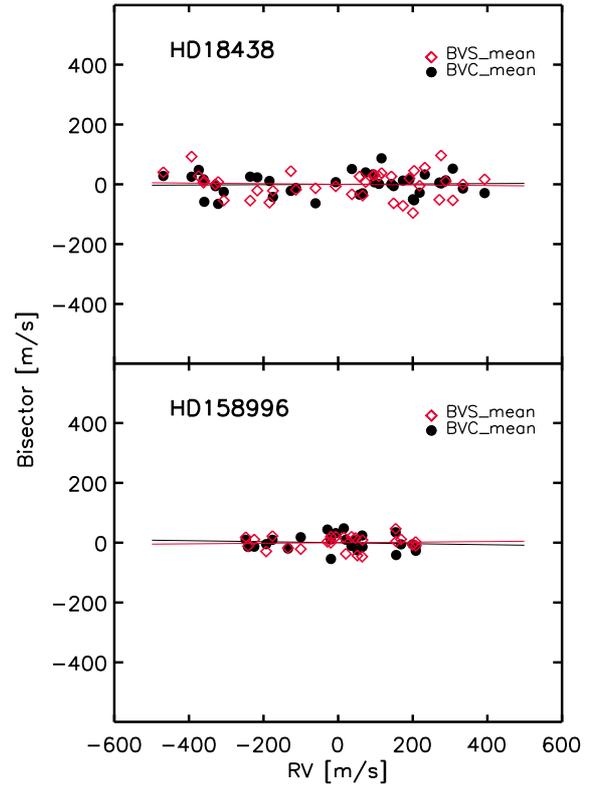}
\caption{RV versus mean values of BVS (open diamonds) and BVC (filled circles) for HD 18438 (top) and HD 158996 (bottom).}.\label{fig:jkasfig10}
\end{figure}

We found reliable RV variations of period of 719.0 days with a FAP of $<$ 10$^{-6}$ (top panel of Figure~\ref{fig:jkasfig5}). BVS and BVC variations show no correlations with RV variations as shown in Figure~\ref{fig:jkasfig10}. On the other hand, we found slight emission features near the center of \ion{Ca}{ii} H lines (Figure~\ref{fig:jkasfig8}), and further checked H$_{\alpha}$ EW variations as another chromospheric indicator (bottom panel of Figure~\ref{fig:jkasfig5}). The periods of $\emph{HIPPARCOS}$ and H$_{\alpha}$ EW variations are close to that of RV variations in the Lomb-Scargle periodogram (Figure~\ref{fig:jkasfig5}).

\begin{figure}[!t]
\centering
\includegraphics[trim=2mm 5mm 8mm 25mm, clip, width=84mm]{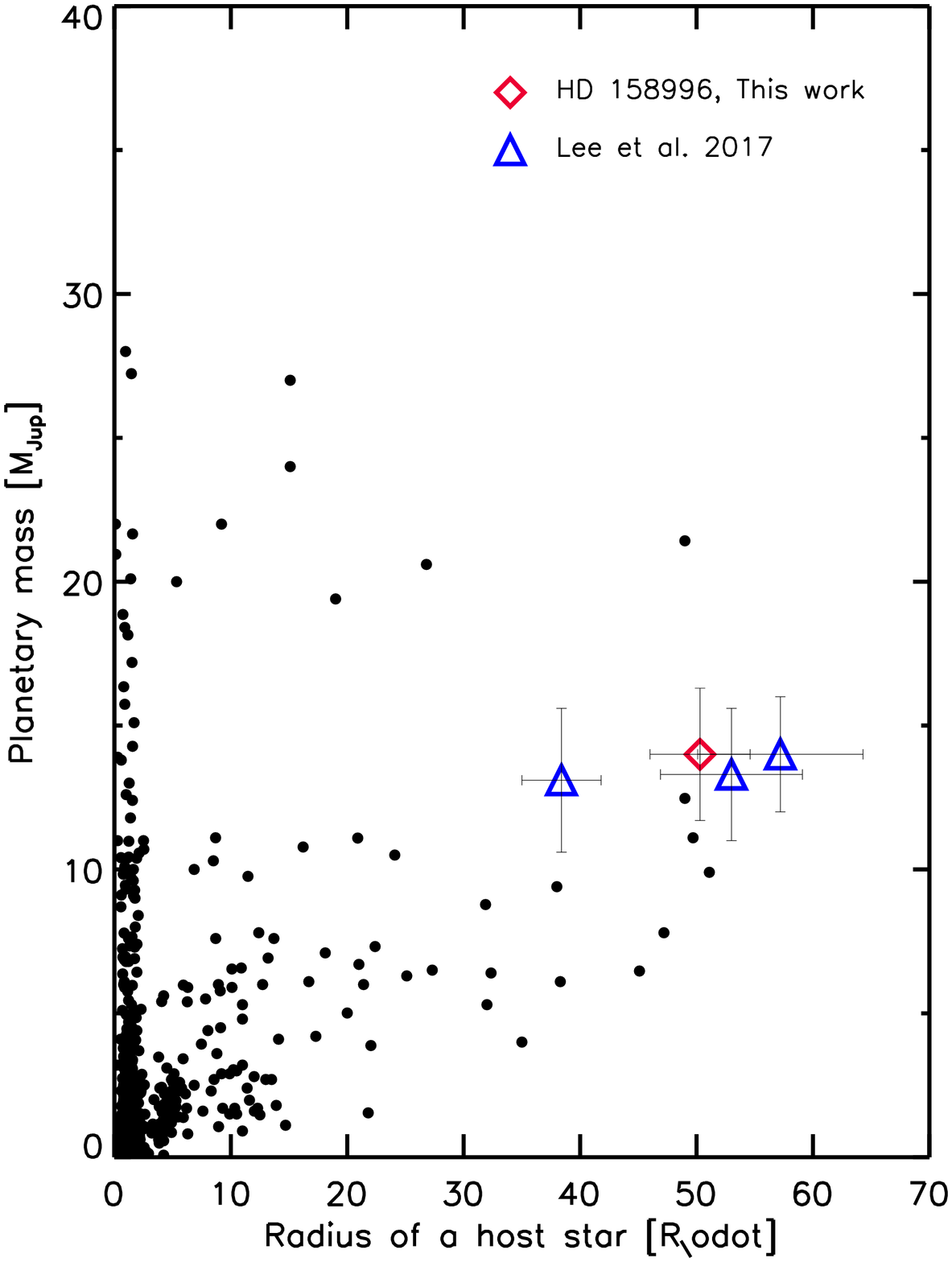}
\caption{Distribution of mass of planetary companions versus stellar radii as of May 2017. Closed circles are known companions and diamond is planetary companion of HD158996 in this work.}.\label{fig:jkasfig11}
\end{figure}

\subsection{HD 158996}

We found reliable RV variations of period of 820.2 days with a FAP of $<$ 10$^{-3}$ (top panel of Figure~\ref{fig:jkasfig6}). BVS and BVC variations show no correlations with RV variations as shown in Figure~\ref{fig:jkasfig10}. \ion{Ca}{ii} H lines do not show meaningful, if any, line reversal at the center. There are no noticable peaks at the level of FAP of $=$ 0.01 in the Lomb-Scargle periodograms (Figure~\ref{fig:jkasfig6}).

\begin{table}[b]
\caption{Orbital parameters for HD 158996 b.}\label{tab:jkastable4}
\centering 
\begin{tabular}{ccc}
\toprule
Parameter & Unit & HD 158996 b\\
\midrule 
P & days & 820.2 $\pm$ 14.0\\
$T_{\rm{periastron}}$& JD &2454993 $\pm$ 128 \\
$K$& m s$^{-1}$ &207 $\pm$ 14\\
$e$ & & 0.13 $\pm$ 0.05 \\
$\omega$ & deg & 168 $\pm$ 12\\
\emph{m}~sin~$\it i$& $\rm{M_{Jup}}$ &14.0 $\pm$ 2.3\\
$\it{a}$ & AU  &2.1 $\pm$ 0.2\\
rms & m s$^{-1}$ & 57.8\\
\bottomrule 
\end{tabular}
\end{table}

\section{Summary and Discussion\label{sec:con}}

We have carried out the SENS program over seven years using the BOES at the 1.8 meter telescope of BOAO in Korea. We observed 38 spectra for HD 18438 from November 2010 to January 2017 and 24 spectra for HD 158996 from June 2010 to January 2017.

We obtained RV measurements, stellar parameters, and RV variation periods from these spectra and found long-period RV variations in HD 18438 and HD 158996. To analyse the origins of the observed long-period RV variations, we checked $\emph{HIPPARCOS}$ photometry, chromospheric activities, and bisectors.

The periods of $\emph{HIPPARCOS}$ photometric and H$_{\alpha}$ EW variations for HD 18438 are close to that of RV variations in Lomb-Scargle periodogram. Since an orbiting companion in general does not produce photometric variations, unless eclipsing, and rotational modulation of surface features induce changes in line bisectors, and the observed RV variation period of 719 days is much longer than the rotational period, we thus conclude that HD 18438 has a RV variation period of 719 days, likely to be caused by pulsations. Periods of fundamental modes of radial pulsation for giants are much shorter than our observed RV variations. Although the exact nature of Long-Secondary Periods (LSP) in giants is still unknown, \cite{Wood} speculated that the origin of long-period RV variations for LSP giants, like HD 18438, can be explained as non-radial pulsation in a low-degree $g^{+}$ mode with star spot activities. Thus we suspect that the origin of observed RV variation period of 719 days for HD 18438 may be related to the this mechanism, operating in LSP giants. The similarity between the periods of RV variations, $\emph{HIPPARCOS}$ photometric, and H$_{\alpha}$ EW variations for HD 18438 and those of LSPs support this conclusion. However, the amplitudes of RV variations for HD 18438 is much smaller than LSPs. More observations and studies regarding the origin of RV variations for HD 18438 are needed.

On the other hand, the observed RV variations period of 820.2 days for HD 158996 are likely to be caused by the planetary companion. The facts that the probability that the real rotational period can be as long or longer than the RV period is very small, periodograms of $\emph{HIPPARCOS}$ photometry and line bisectors do not show any peaks at the RV period, no variations of \ion{Ca}{ii} H line profile at different phases are seen, line bisectors do not show any correlation with RV, and RV variations are well fitted by Keplerian variation lead us to conclude that observed RV variations are very likely to be the result of orbital motion. Assuming a mass of 1.8 $M_{\odot}$ for HD 158996, we calculated parameters of planetary companion candidate HD 158996 b, the minimum mass of 14.0 $\rm{M_{Jup}}$, the semi-major axis of 2.1 AU and eccentricity ($\it{e}$) of 0.13. The orbital parameters for HD 158996 b are shown in Table~\ref{tab:jkastable4}.

Figure~\ref{fig:jkasfig11} shows the distribution of mass of planetary companions versus stellar radii. Triangles are planetary companion candidates of SENS program (\citealt{Lee17}). This study shows that HD 158996 (diamond) is the brightest star harbouring planetary companion so far. Further study of this extreme giant star will help us understand how evolved stars affect exoplanets, as previously described (\citealt{Villaver}; \citealt{Mustill}). HD 18438 also has a very large radius that will help us understand the origins of long-period RV variations of giants. Our results can be used for subsequent planetary and stellar research based on the characteristics of the host star.


\acknowledgments

We thank the anonymous referee for useful comments. BCL acknowledges partial support by Korea Astronomy and Space Science Institute (KASI) grant 2017-1-830-03. Support for MGP was provided by the KASI under the R\&D program supervised by the Ministry of Science, ICT and Future Planning and by the National Research Foundation of Korea to the Center for Galaxy Evolution Research (No. 2017R1A5A1070354). This work was supported by BK21 Plus of National Research Foundation of Korea.


\end{document}